\RequirePackage{lineno}
\documentclass[aps,prl,twocolumn,superscriptaddress,showpacs]{revtex4}

\usepackage{epsfig}
\usepackage{amsmath,amssymb}
\usepackage{graphicx,color}
\usepackage{dcolumn}
\usepackage{bm}

\usepackage{ifpdf}
\ifpdf
\DeclareGraphicsExtensions{.pdf, .jpg, .tif}
\usepackage[%
  pdftitle={Reevaluation of the parton distribution of strange quarks in the nucleon},%
  pdfauthor={The HERMES Collaboration},%
  pdfsubject={HERMES Strangeness},%
  pdfstartview=FitH,%
  bookmarks=true,%
  bookmarksopen=true,%
  breaklinks=true,%
  colorlinks=true,%
  linkcolor=blue,anchorcolor=blue,%
  citecolor=blue,filecolor=blue,%
  menucolor=blue,pagecolor=blue,%
  urlcolor=blue]{hyperref}
\else
\DeclareGraphicsExtensions{.eps, .jpg}
\usepackage[%
  breaklinks=true,%
  colorlinks=true,%
  linkcolor=blue,anchorcolor=blue,%
  citecolor=blue,filecolor=blue,%
  menucolor=blue,pagecolor=blue,%
  urlcolor=blue]{hyperref}
\fi

\begin{document}



\title{Transverse polarization of $\Lambda$ hyperons from quasireal photoproduction on nuclei}

\def\hermes{{\sc Hermes}}
\def\clas{{\sc Clas}}
\def\jlab{{\sc JLab}}
\def\desy{{\sc Desy}}
\def\hera{{\sc Hera}}
\def\cornell{{\sc Cornell}}
\def\cea{{\sc Cea}}
\def\lepto{{\sc Lepto}}
\def\pepsi{{\sc Pepsi}}
\def\jetset{{\sc Jetset}}
\def\pythia6{{\sc Pythia6}}
\def\pythia{{\sc Pythia}}
\def\radgen{{\sc Radgen}}
\def\geant{{\sc Geant}}


\def\groupargonne{\affiliation{Physics Division, Argonne National Laboratory, Argonne, Illinois 60439-4843, USA}}
\def\groupbari{\affiliation{Istituto Nazionale di Fisica Nucleare, Sezione di Bari, 70124 Bari, Italy}}
\def\groupbeijing{\affiliation{School of Physics, Peking University, Beijing 100871, China}}
\def\groupbilbao{\affiliation{Department of Theoretical Physics, University of the Basque Country UPV/EHU, 48080 Bilbao, Spain and IKERBASQUE, Basque Foundation for Science, 48013 Bilbao, Spain}}
\def\groupcolorado{\affiliation{Nuclear Physics Laboratory, University of Colorado, Boulder, Colorado 80309-0390, USA}}
\def\groupdesy{\affiliation{DESY, 22603 Hamburg, Germany}}
\def\groupzeuthen{\affiliation{DESY, 15738 Zeuthen, Germany}}
\def\groupdubna{\affiliation{Joint Institute for Nuclear Research, 141980 Dubna, Russia}}
\def\grouperlangen{\affiliation{Physikalisches Institut, Universit\"at Erlangen-N\"urnberg, 91058 Erlangen, Germany}}
\def\groupferrara{\affiliation{Istituto Nazionale di Fisica Nucleare, Sezione di Ferrara and Dipartimento di Fisica e Scienze della Terra, Universit\`a di Ferrara, 44122 Ferrara, Italy}}
\def\groupfrascati{\affiliation{Istituto Nazionale di Fisica Nucleare, Laboratori Nazionali di Frascati, 00044 Frascati, Italy}}
\def\groupgent{\affiliation{Department of Physics and Astronomy, Ghent University, 9000 Gent, Belgium}}
\def\groupgiessen{\affiliation{II. Physikalisches Institut, Justus-Liebig Universit\"at Gie{\ss}en, 35392 Gie{\ss}en, Germany}}
\def\groupglasgow{\affiliation{SUPA, School of Physics and Astronomy, University of Glasgow, Glasgow G12 8QQ, United Kingdom}}
\def\groupillinois{\affiliation{Department of Physics, University of Illinois, Urbana, Illinois 61801-3080, USA}}
\def\groupmichigan{\affiliation{Randall Laboratory of Physics, University of Michigan, Ann Arbor, Michigan 48109-1040, USA}}
\def\groupmoscow{\affiliation{Lebedev Physical Institute, 117924 Moscow, Russia}}
\def\groupnikhef{\affiliation{National Institute for Subatomic Physics (Nikhef), 1009 DB Amsterdam, The Netherlands}}
\def\groupstpetersburg{\affiliation{B.P. Konstantinov Petersburg Nuclear Physics Institute, Gatchina, 188300 Leningrad region, Russia}}
\def\groupprotvino{\affiliation{Institute for High Energy Physics, Protvino, 142281 Moscow region, Russia}}
\def\groupregensburg{\affiliation{Institut f\"ur Theoretische Physik, Universit\"at Regensburg, 93040 Regensburg, Germany}}
\def\grouprome{\affiliation{Istituto Nazionale di Fisica Nucleare, Sezione di Roma, Gruppo Collegato Sanit\`a and Istituto Superiore di Sanit\`a, 00161 Roma, Italy}}
\def\grouptriumf{\affiliation{TRIUMF, Vancouver, British Columbia V6T 2A3, Canada}}
\def\grouptokyo{\affiliation{Department of Physics, Tokyo Institute of Technology, Tokyo 152, Japan}}
\def\groupamsterdam{\affiliation{Department of Physics and Astronomy, VU University, 1081 HV Amsterdam, The Netherlands}}
\def\groupwarsaw{\affiliation{National Centre for Nuclear Research, 00-689 Warsaw, Poland}}
\def\groupyerevan{\affiliation{Yerevan Physics Institute, 375036 Yerevan, Armenia}}
\def\groupnone{\noaffiliation}


\groupargonne
\groupbari
\groupbeijing
\groupbilbao
\groupcolorado
\groupdesy
\groupzeuthen
\groupdubna
\grouperlangen
\groupferrara
\groupfrascati
\groupgent
\groupgiessen
\groupglasgow
\groupillinois
\groupmichigan
\groupmoscow
\groupnikhef
\groupstpetersburg
\groupprotvino
\groupregensburg
\grouprome
\grouptriumf
\grouptokyo
\groupamsterdam
\groupwarsaw
\groupyerevan


\author{A.~Airapetian}  \groupgiessen \groupmichigan
\author{N.~Akopov}  \groupyerevan
\author{Z.~Akopov}  \groupdesy
\author{E.C.~Aschenauer}  \groupzeuthen 
\author{W.~Augustyniak}  \groupwarsaw
\author{R.~Avakian}  \groupyerevan
\author{A.~Avetissian}  \groupyerevan
\author{E.~Avetisyan}  \groupdesy
\author{S.~Belostotski}  \groupstpetersburg
\author{N.~Bianchi}  \groupfrascati
% ...

\collaboration{The HERMES Collaboration} \noaffiliation

\date{\today \ -- final version }

\begin{abstract}
The transverse 
polarization of $\Lambda$ hyperons was measured in inclusive quasireal photoproduction for various target nuclei ranging from hydrogen to xenon. 
The data were obtained by the \hermes\ experiment at \hera\ 
using the 27.6~GeV lepton beam 
and nuclear gas targets internal to the lepton storage ring. The 
polarization observed 
is positive for light target nuclei and is
compatible with zero for 
krypton and xenon.
\end{abstract}

\maketitle


\section{ Introduction}

The transverse polarization
of ${\Lambda}$ hyperons 
produced in inclusive unpolarized hadron-nucleon, hadron-nucleus, and nucleus-nucleus collisions 
at high energies is a well-established phenomenon. 
The polarization $P_{\rm n}^{\Lambda}$ measured in these experiments with hadron beams 
is perpendicular to the production plane spanned by the momentum vectors $\vec{k}$ and 
$\vec{p}$ of the beam and the 
produced $\Lambda$, respectively, which is the only direction allowed by parity 
conservation for an axial-vector quantity and unpolarized beams and targets .
A substantial transverse $\Lambda$ 
polarization was first observed in proton-beryllium 
collisions at a proton-beam energy of 300~GeV \cite{Bunce}. 
Since then, numerous experiments using a variety of beams and targets 
have been performed to study this effect in detail. 
For a review of experiments and results see, e.g., Refs.~\cite{Pondrom,Apostolos,Abramov}.  
The polarization is essentially independent of the beam momentum; its magnitude  
rises with $p_T$, the $\Lambda$ momentum component transverse 
to the beam direction, for $p_T$ 
values up to about 1 GeV, where it reaches values of up to $|P_{\rm n}^{\Lambda}| \approx 0.4$;
then, it is independent of $p_T$ up to the highest measured $p_T$ values 
of about 3 GeV.
At fixed $p_T$, 
{\bf $|P_{\rm n}^{\Lambda}|$} rises with the Feynman variable 
$x_F = p^*_L/p^*_{L,max}$, where $p^*_L$ 
is the component of the $\Lambda$  momentum in the beam direction measured in the beam-target 
center-of-mass system and $p^*_{L,max}$ is its maximal possible value.  
The transverse polarization depends only weakly on the atomic-mass number $A$ of the target 
nuclei~\cite{Pondrom,Apostolos,Castillo}, whereas the polarization 
for beryllium appeared to be of slightly 
smaller magnitude than that for H and D~\cite{Raychaudhuri}. A relative reduction of 
the magnitude of the polarization by about 
20$\%$ was observed
for copper and lead targets compared to beryllium.
Furthermore, $P_{\rm n}^{\Lambda}$ 
remained small in relativistic heavy-ion collisions up to $p_T \approx 2.5$~GeV. At higher 
transverse momenta, $P_{\rm n}^{\Lambda}$ measured in such experiments was found to be similar 
to that observed in p-p or p-nucleus scattering~\cite{Bellwied}. 
For proton and neutron beams, the measured polarization is negative, 
i.e., the polarization vector $\vec{P}_{\rm n}^{~\Lambda}$ and the normal 
$\widehat{n} = {\vec{k} \times \vec{p} }/{|\vec{k} \times \vec{p}|}$ to the production plane have opposite directions, 
while for $K^-$~\cite{ganguli,gourlay,haupt} and 
$\Sigma^-$~\cite{wa89,selex} beams the polarization is positive in the forward direction ($x_F>0$) and for $\pi^-$ beams~\cite{stuntebeck,sughara,bensinger} it is positive in the backward hemisphere ($x_F<0$). In contrast, the polarization is compatible with zero for $\pi^+$ and $K^+$ beams.

While the transverse  polarization of hyperons was studied in detail with hadron beams, very 
little experimental information about $P_{\rm n}^{\Lambda}$ is available from photo- or 
electroproduction. 
The $x_F$ dependence observed in early measurements~\cite{Aston,Abe} 
was very similar to the one seen in $\pi^-$-p reactions, but the data  
were of low statistical accuracy. 
More recently, the \hermes\ experiment has obtained for the  
first time statistically significant experimental results on the transverse $\Lambda$ 
polarization in inclusive quasireal photoproduction~\cite{Airapetian}. The measured
polarization is positive in both the forward and the backward directions.

The analysis presented in Ref.~\cite{Airapetian} combined the data collected by the \hermes\ 
experiment in the years 1996--2000 using mostly hydrogen and deuterium targets. More data were 
collected in the years 2002--2005 with the target nuclei 
H, D, $^4$He, Ne, Kr, and Xe,
which allowed for the study of the dependence of
$P_{\rm n}^{\Lambda}$ on the atomic-mass number $A$ of the target nuclei. The results of all 
these measurements are presented here.

\section{Experiment} 

The data were accumulated by the \hermes\ experiment at the \hera\ accelerator facility of 
\desy. The 27.6~GeV lepton (electron or positron) beam passed through a
40~cm long open-ended tubular 
storage cell internal to the  lepton storage ring. 
The storage cell was filled with polarized or unpolarized target gas of the various elements. 
Part of the data was collected using a transversely or longitudinally
polarized hydrogen target and a longitudinally polarized deuterium target, respectively. The 
direction of the target 
polarization was reversed  in 1--3~min intervals, resulting in a vanishing average 
target polarization of the data set.

The \hermes\ detector is described in detail in Ref.~\cite{Ackerstaff}. It was a forward 
magnetic spectrometer with a geometric acceptance confined to two regions in scattering angle, 
arranged symmetrically above (top) and below (bottom) the  
beam pipe and covering ranges of $\pm$(40--140)~mrad in the vertical and $\pm$170~mrad in the 
horizontal component of the scattering angle with respect to the center of the target cell.

The criteria for data selection and the analysis procedure are similar to those described in 
detail in Ref.~\cite{Airapetian}. 
The scattered lepton was not required to be detected. In this case the data sample is dominated 
by the kinematic regime  of quasireal photoproduction with 
$Q^2 \approx 0$~GeV$^2$, (where $-Q^2$ represents the squared four-momentum of the virtual 
photon exchanged in the electromagnetic interaction).  
The kinematic distribution of the quasireal photons  was obtained from a Monte Carlo simulation of the process using the 
\pythia\ event generator~\cite{PYTHIA} 
and a \geant\ \cite{GEANT} model of the detector.  
The energy distribution obtained from this simulation is nearly flat in the region $\nu \approx$ 6--26~GeV with a broad maximum near 13~GeV. 
Below and above this region it drops rapidly to zero, resulting in an average photon energy of
$\langle \nu \rangle \approx 16$~GeV.

The $\Lambda$ events were detected through their $\Lambda \to \text{p}\pi ^{-}$ decay channel, by 
requiring the presence of at least two hadron candidates of opposite charges. 
In the event selection the fact was used that in the laboratory system  the momentum of the 
decay proton is always larger than the pion momentum for  $\Lambda$ momenta above  $\sim 300$~MeV.
The information from a 
dual-radiator ring-imaging Cherenkov counter was used to assure that the positive hadron was not a pion.
When more than one positive or negative hadron was found in an event, all possible combinations 
of oppositely charged hadron pairs were considered. 
All tracks were also required to satisfy a series of fiducial-volume cuts designed to avoid the 
inactive edges of the detector. Furthermore, both hadron tracks were required to be 
reconstructed in the same spectrometer half to avoid effects caused by  possible misalignment of the 
two spectrometer halves relative to each other. This requirement reduced the number of 
$\Lambda$-event candidates by $\sim 15\%$.


\begin{figure}
\epsfig{width=0.9\columnwidth, file=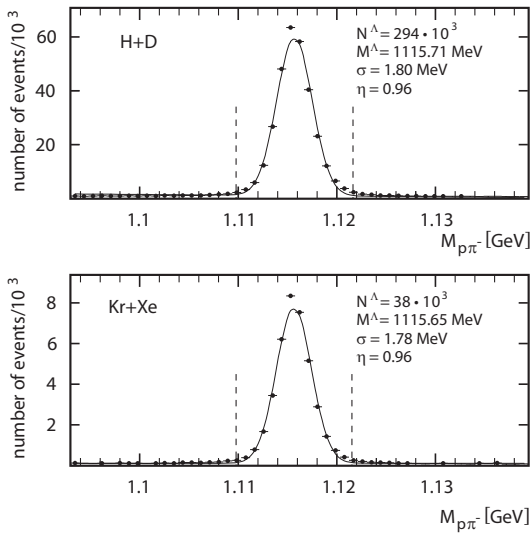}
\caption{Invariant-mass distributions for $\Lambda$ events obtained with hydrogen and deuterium 
targets (top panel) and with krypton and xenon targets (bottom panel). The vertical 
lines indicate the invariant-mass interval used for the determination of
the $\Lambda$ polarization. The quantities given in the legends are the number of analyzed $\Lambda$ events, ${N^\Lambda}$, in the selected 
invariant-mass window after subtraction of background events, the reconstructed $\Lambda$ mass ${\rm M}^{\Lambda}$, the resolution $\sigma$ of the invariant-mass 
distribution, and the fraction $\eta$  of 
$\Lambda$ events in this mass window. 
}
\label{fig:mass}
\end{figure}


Track reconstruction was performed with a track-fitting algorithm based on the Kalman filter with 
substantially improved vertex determination and momentum resolution compared to the one used for the data 
published previously. This algorithm allows for the  
best-possible estimates on track parameters at the beam crossing and/or at 
the (possible) vertices with other tracks of a given event.
Two spatial vertices were reconstructed for each event: the $\Lambda$-decay vertex from the 
intersection of the proton and pion tracks and the $\Lambda$-production vertex from the 
intersection of the reconstructed  $\Lambda$ track with the nominal beam axis.
The $\Lambda$-production vertex was required to be downstream of the upstream end of the 
target cell and the decay vertex was required to be at least 40 cm downstream of the center of the 
target cell. The latter requirement was chosen as a compromise between statistical precision and low background of the data 
sample and the need to avoid---for data taken with a polarized target---any residual influence of the 
target's magnetic field on the polarization measurement.

For tracks fulfilling all the above given requirements the invariant mass of the hadron pair was evaluated. 
The resulting invariant-mass distributions for the combined hydrogen and deuterium (H+D) and 
the combined krypton and xenon (Kr+Xe) data are shown in Fig.~\ref{fig:mass}. These 
distributions are very similar for all nuclei. They were fitted by a Gaussian plus a second-order polynomial line shape. 
Compared to the data published previously~\cite{Airapetian}, the resolution $\sigma$ of the   
mass reconstruction was improved from 2.23 MeV to 1.80 MeV. The position 
of the $\Lambda$ peak agrees within $\sim0.10$ MeV  
with the world average of (1115.683 $\pm$ 0.006) MeV~\cite{PDG}. Events within an 
invariant-mass window 
of ${\pm 3.3 \sigma}$ around the mean value of the Gaussian fit were selected, and background 
events, $N^{\rm bgr}$, were subtracted with the procedure
described in Ref.~\cite{Airapetian}. The background is small, the fraction 
$\eta = N^{\Lambda}/(N^{\Lambda} + N^{\rm bgr})$ of $\Lambda$ events in the selected mass
window being $\sim 96\%$ for all targets. 

\section{Extraction of ${\Lambda}$~polarization}

{
\begin{figure}
\epsfig{width=0.8\columnwidth, file=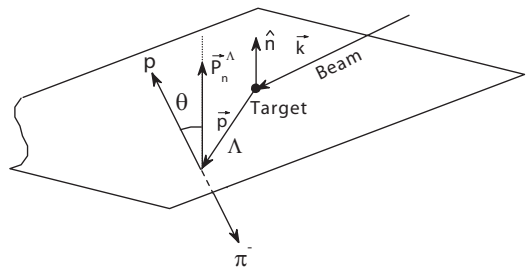}
\caption{Sketch of $\Lambda$ production and decay. The polarization vector 
$\vec{P}_{\rm n}^{\Lambda}$ is directed along 
the normal $\widehat{n}$ to the $\Lambda$ production plane; $\theta$ is the angle between the momentum of the decay proton  and 
$\widehat{n}$ in the rest frame of the $\Lambda$ hyperon.
}
\label{fig:diag}
\end{figure}
}

The topology of $\Lambda$ production and decay is sketched in Fig.~\ref{fig:diag} where the 
decay into a proton and a pion is shown in the $\Lambda$ rest frame. The method of extraction of the transverse $\Lambda$ polarization is described in detail in 
Ref.~\cite{Airapetian}. For the parity-violating $\Lambda \to \text{p}\pi ^{-}$ decay, the angular 
distribution of the decay protons in the $\Lambda$ rest frame is given by
\begin{equation}
\frac{dN}{d\Omega}=\frac{dN_0}{d\Omega}(1+\alpha P_{\rm n}^\Lambda\cos\theta),
\label{eq:eq2}
\end{equation}

\noindent where $\frac{dN_{0} }{d\Omega  }$ is the decay distribution for unpolarized $\Lambda$ 
hyperons, $\theta $ is the angle between the proton momentum and the normal $\widehat{n}$,
and $\alpha=0.642 \pm 0.013$~\cite{PDG}. 
The decay 
protons are preferentially emitted along the spin direction of the $\Lambda$ in its 
rest frame. This provides 
the possibility of obtaining the $\Lambda$ polarization by measuring the asymmetry in the proton's 
angular distribution. 

For a detector with $4\pi$ acceptance, the polarization is given by 
$P_{\rm n}^\Lambda=\frac{3}{\alpha}\left\langle\cos\theta\right\rangle$, where 
$\left\langle\cos\theta\right\rangle\equiv\frac{1}{N^{\Lambda}}
\sum_{i=1}^{N^{\Lambda} }(\cos \theta)_i$ is the first moment of the angular distribution and 
$N^{\Lambda}$ is the number of $\Lambda$ events analyzed.
For a detector with nonuniform acceptance, the linear $\cos\theta$ distribution in Eq.~\ref
{eq:eq2} can be strongly distorted. For the determination of $P_{\rm n}^{\Lambda}$ the mean 
value of $\cos\theta$ for the \textit{unpolarized }distribution, 
$\left\langle\cos\theta\right\rangle_0$, must be known with 
good precision. 
For a detector with ideal top/bottom mirror symmetry, 
there is an important 
simplification: the first moment for unpolarized $\Lambda$ events vanishes ($\left\langle \cos 
\theta  \right\rangle_0 =0$), and 
$\left\langle \cos^m\theta  \right\rangle_0 = \left\langle \cos^m\theta  \right\rangle$ 
($ m = 2, 4, ...$), i.e., 
 all even ``polarized'' moments are equal to the ``unpolarized'' ones. This allows for
the determination of $P_{\rm n}^{\Lambda}$ 
using only the experimentally measured values 
for $(\cos \theta )_i$ without the need for a  Monte Carlo simulation of the spectrometer 
acceptance:
\begin{equation}
P_{\rm n}^\Lambda =\frac{\left\langle \cos \theta  \right\rangle}
{\alpha \left\langle \cos ^{2} \theta  \right\rangle} .
\label{eq:eq3}
\end{equation}

The numerator in Eq.~\ref{eq:eq3} represents the measured asymmetry in the angular distribution 
while the denominator stands for the analyzing power in the case of nonuniform acceptance. 
For the \hermes\ spectrometer, the top/bottom mirror symmetry is not absolutely perfect and
Eq.~\ref{eq:eq3} is only a good approximation. The true values for 
$\left\langle\cos\theta\right\rangle_0$ and $P_{\rm n}^{\Lambda}$  therefore have to be 
determined in an iterative procedure that  
takes into account the measured differences between $\langle\cos\theta\rangle$ for the top and 
bottom halves of the detector~\cite{Airapetian}.

\section{Results}

\begin{table}[t]
\caption{Average transverse ${\Lambda}$ polarization $P_{\rm n}^{\Lambda}$ for various target 
nuclei together with the statistical uncertainties $\delta{P_{\rm n}^{\Lambda}}({\rm stat})$ 
of the measurements, the number of analyzed $\Lambda$ events, ${N^\Lambda}$, in the selected 
invariant-mass window after subtraction of background events, and the fraction $\eta$  of 
$\Lambda$ events in this mass window. Also presented are the differences 
$\Delta{\rm M}^{\Lambda}$ between the reconstructed $\Lambda$ mass and the world 
average~\cite{PDG},
the resolution $\sigma$ of the invariant-mass 
distribution, and the average values of  the transverse $\Lambda$ momentum 
$\left\langle p_T \right\rangle $ and the 
variable $\left\langle \zeta \right\rangle $, which is an approximate measure of whether the 
$\Lambda$ is produced in the forward or in the backward region (see text). The systematic uncertainty is 0.02 for all targets.}
\label{tab:results}
\vspace{3mm}
\begin{tabular}{lrrrrrr}
\hline\hline\\[-2mm]
 & ~~H~~~ & ~~D~~~  & ~~$^4$He~~ & ~~Ne~~~ & ~~~Kr~~~ & ~~Xe~~~ \\
\hline\\[-2mm] 
$P_{\rm n}^{\Lambda}$ & 0.062& 0.052& 0.051 & 0.092& -0.005& 0.010\\[1mm] 
$\delta{P_{\rm n}^{\Lambda}}({\rm stat})$ & 0.008&0.006 & 0.044& 0.026& 0.017& 0.023\\[1mm] 
${N^\Lambda}/10^3$ &108.5 &185.9 & 3.4 & 10.2 & 24.2& 13.7\\[1mm]  
$\eta$ &0.96 &0.96 & 0.96&0.96 &0.96 & 0.97\\[1mm]  
$\Delta{\rm M}^{\Lambda}$ [MeV] &0.02 &0.05 & 0.09&0.11 & 0.04& 0.00\\[1mm]  
$\sigma$ [MeV] & 1.79&1.82 & 1.96& 1.89&1.77 & 1.79\\[1mm]  
$\left\langle p_T \right\rangle $ [GeV]~~~ &0.63 & 0.63& 0.67&0.68 &0.64 & 0.64\\[1mm]  
$\left\langle \zeta \right\rangle $  & 0.25& 0.25& 0.27& 0.27&0.25 &0.25 \\[1mm]  
\hline\hline
\end{tabular}
\end{table}

The experimental results for the extracted transverse polarization $P_{\rm n}^{\Lambda}$, 
averaged over 
all kinematic variables, are presented in Table~\ref{tab:results} for the various target nuclei, 
together with the statistical  
uncertainties of the measurements, 
the number of $\Lambda$ events in the selected
invariant-mass window after subtraction of background events, and
the fraction $\eta$  of $\Lambda$ events in this mass 
window.  Also presented are the differences $\Delta{\rm M}^{\Lambda}$ between the reconstructed 
$\Lambda$ mass and the world average~\cite{PDG},
the resolution $\sigma$ of the invariant-mass distribution, the
average values of the transverse $\Lambda$ momentum, $p_T$, and of the variable 
$\zeta =(E^{\Lambda}+p_L)/({E+k}) \approx E^{\Lambda} /E $. Here $E^{\Lambda}$ and $E$ are the 
energies of the $\Lambda$ produced and the beam lepton, respectively, and $p_L$ is the 
$\Lambda$'s momentum component in the beam direction measured in the target rest frame. 
As discussed in Ref.~\cite{Airapetian}, the variable $\zeta$
provides an approximate measure of whether
the $\Lambda$ hyperon was produced in the forward region $(\zeta >0.3)$ in the center-of-mass frame of the $\gamma^*$-nucleon reaction, 
whereas for $\zeta<0.2$ it is predominantly produced
in the backward region but still with a significant admixture of forward-going $\Lambda$ 
hyperons. 
The values of $\Delta{\rm M}^{\Lambda}$, $\sigma$, $\left\langle p_T \right\rangle $, and $\left\langle \zeta \right\rangle $
are very similar for all targets.

The systematic uncertainty of $P_{\rm n}^{\Lambda}$ has been estimated to be $\pm 0.02$. This 
value was 
derived from detailed Monte Carlo studies that took into account possible detector 
misalignments, 
and also from the false polarization measured for
$K^0_s \rightarrow \pi^+ \pi^-$ events that provide an event topology with two separated vertices
similar to ${\Lambda}$ decays. Some data were taken with a transversely polarized hydrogen target. The effects of the 
transversely oriented magnetic holding field in the target region were taken into account in the 
reconstruction of the charged-particle tracks. The integrated transverse target field was
$\sim 0.17$~Tm. This caused an average precession of the $\Lambda$'s magnetic moment of less than
one degree, with negligible impact on the extracted transverse polarization.

\begin{figure}[t]
\begin{center}
\epsfig{width=0.90\columnwidth, file=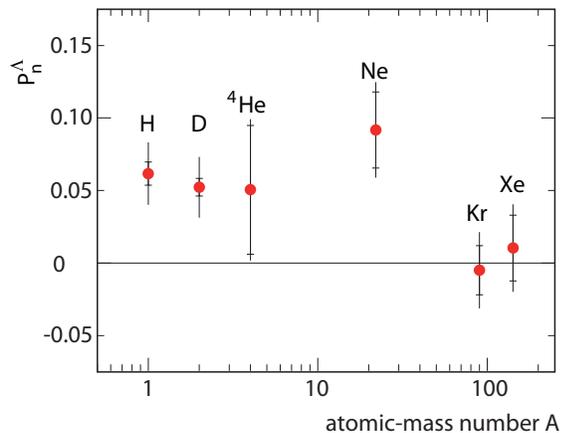}
\end{center}
\caption{Dependence of the transverse polarization $P_{\rm n}^{\Lambda}$ 
on the atomic-mass number $A$ of the target nuclei. The inner error bars represent the 
statistical uncertainties; the full error bars represent the total uncertainties, evaluated as 
the sum in quadrature of statistical and systematic uncertainties. 
}
\label{fig:adep}
\end{figure}

The results for the measured transverse polarizations are presented in Fig.~\ref{fig:adep} as a 
function of the atomic-mass number $A$ of the various target nuclei. The results for the light 
nuclei are significantly positive; those for hydrogen and deuterium agree within their 
statistical uncertainties. The average value for H+D is 
$\left\langle P_{\rm n}^{\Lambda}({\rm H+D})\right\rangle = 0.056 \pm 0.005\text{(stat)} \pm 0.020\text{(sys)}$ 
with the average value for all nuclei being $\left\langle P_{\rm n}^{\Lambda}(\text{all }A)\right\rangle = 0.044 \pm 0.011$.

The transverse polarization for neon is above this value by more than one standard deviation, 
while  the results  for krypton and xenon are 
compatible with zero within the statistical uncertainties of the data. The average value of 
$P_{\rm n}^{\Lambda}$ for the combined krypton and xenon data is 
$\left\langle P_{\rm n}^{\Lambda}({\rm Kr+Xe})\right\rangle = 0.000 \pm 0.014\text{(stat)} \pm  0.020\text{(sys)}$.

Despite the rather large value for 
neon there is an indication of a decrease of $P_{\rm n}^{\Lambda}$ with the atomic-mass number 
$A$ of the target nuclei. However, the statistical accuracy of the measurements does not allow for
a precise determination of the functional form of this $A$ dependence.

\begin{figure}[t]
\begin{center}
\epsfig{width=0.75\columnwidth, file=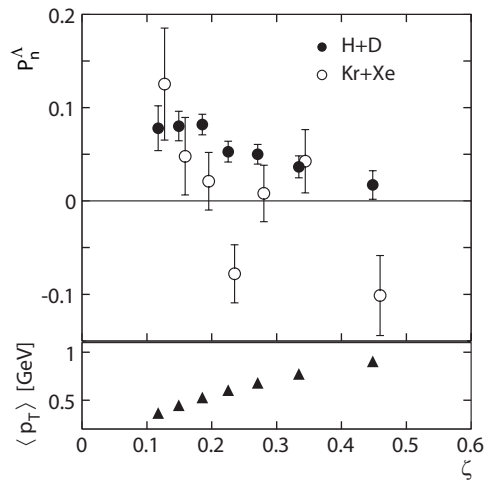}
\end{center}
\caption{Dependence of the transverse polarization $P_{\rm n}^{\Lambda}$ 
for the combined hydrogen and deuterium data (closed symbols) and the combined krypton and 
xenon data (open symbols) on the variable $\zeta$. The error bars represent statistical uncertainties only; the systematic uncertainties
are not shown, since they are strongly correlated for the kinematic dependences. The values of $\left\langle p_T 
\right\rangle$ for each $\zeta$ bin are shown in the lower panel. 
}
\label{fig:zetadep}
\end{figure}

The $\Lambda$ polarizations for the combined H+D and the combined Kr+Xe data are shown as a 
function of $\zeta$ in Fig.~\ref{fig:zetadep}.
The H+D data (closed symbols) decrease continuously from a value of $\sim 0.08$ at low $\zeta$ 
to $\sim 0.02$ at $\zeta \simeq 0.45$, while the Kr+Xe data (open symbols) fluctuate around 
zero. For each point in $\zeta$ 
the average value of $p_T$ is different, as it is shown in the lower panel of the figure. 

\begin{figure}[bt]
\begin{center}
\epsfig{width=0.83\columnwidth, file=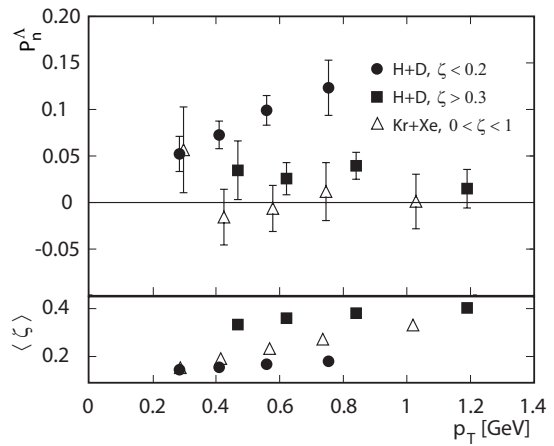}
\end{center}
\caption{Dependence of the transverse polarization $P_{\rm n}^{\Lambda}$ on the transverse $\Lambda$ momentum $p_T$. Closed circles (squares) represent the combined hydrogen and deuterium data for the region $\zeta < 0.2$ ($\zeta > 0.3$). 
The combined krypton and xenon data (open triangles) are shown for the full $\zeta$ range.
 The error bars represent the statistical 
uncertainty.
The values of $\left\langle \zeta \right\rangle$ for each $p_T$ bin are shown in the lower 
panel. 
}
\label{fig:ptdep}
\end{figure}

In Fig.~\ref{fig:ptdep} the polarizations are shown  as a function of $p_T$. The H+D data
are presented for two intervals in the variable $\zeta$. The $p_T$ dependence in these two 
intervals is rather different. In the region $\zeta < 0.2$, where the produced $\Lambda$ 
hyperons mainly stem from the
backward region, the polarization increases linearly with $p_T$ up to a value
of $\sim 0.12$ at $p_T \simeq 0.75$ GeV (closed circles), while in the region 
$\zeta > 0.3$ (closed squares) the polarization is substantially smaller with very
little dependence on $p_T$. The statistical uncertainties of the Kr+Xe data prevent a firm conclusion about the $p_T$ dependences in the two $\zeta$ regions. The polarization is compatible with zero over 
the whole $p_T$ range although the average polarization in the region $\zeta < 0.2$ is $0.059 \pm 0.024 ({\rm stat})$, while it is $-0.012 \pm 0.027 ({\rm stat})$ in the region $\zeta > 0.3$. 
It should be noted that the measured ratio of $\Lambda$ yields for (Kr+Xe) and D decreases with $\zeta$ and increases at large $p_T$. This behavior is rather similar to the ratio of hadron multiplicities for heavy nuclear targets and deuterium as a function of $z$ and $p_t$ in semi-inclusive deep-inelastic scattering~\cite{airapetian07,airapetian11}, where 
$z = E_h/\nu$ is the fractional hadron energy and $p_t$ is the transverse hadron momentum with respect to the virtual-photon direction.

\section{Discussion}
The transverse $\Lambda$ polarization $P_{\rm n}^{\Lambda}$ was measured in inclusive quasireal 
photoproduction on nuclei. The observed polarization is positive, the same as those
observed for 
$K^-$ and $\Sigma^-$ beams in the forward direction and for $\pi^-$ beams in the backward direction.
The polarizations obtained for 
hydrogen and deuterium targets, and consequently for free protons and neutrons, agree within their statistical accuracies.
For the combined hydrogen and deuterium data the measured polarization decreases with the 
variable $\zeta$ that provides an approximate measure of whether
the $\Lambda$ hyperon was produced in the forward or in the backward region in the 
center-of-mass frame of the $\gamma^*$-nucleon reaction. For small $\zeta$ ($\zeta < 0.2$), the 
polarization increases linearly with $p_T$, whereas in the forward region ($\zeta > 0.3$) the 
polarization is substantially smaller with very
little dependence on $p_T$. This behavior points to a different production mechanism
in the two kinematic regions. The interpretation of these 
observations depends on the mechanism assumed for generating the $\Lambda$ polarization,
which at present is not understood. 
There is an indication of a decrease of $P_{\rm n}^{\Lambda}$ with the atomic-mass number $A$ 
of the target nuclei. In contrast to measurements in hadron-nucleus scattering---where the 
magnitude of the polarization for heavier nuclei appears to be somewhat smaller than for light nuclei but still 
substantially different from zero---the polarization for the combined Kr+Xe data from the 
present measurement, integrated over $p_T$ and $\zeta$,
is compatible with zero within the statistical uncertainties and about four standard deviations below the combined H+D result.
At low $\zeta$ and $p_T$, i.e., in the backward hemisphere, the Kr+Xe results seem to agree within their
statistical uncertainties with the H+D results, such that the overall reduction possibly stems mainly from the forward direction.

The situation is further complicated by the fact that
a fraction of the detected $\Lambda$ particles and their polarization originates from decays of 
heavier hyperons like $\Sigma^0$, $\Sigma(1385)$, and $\Xi$.  So far, the transverse 
polarization of these hyperons in quasireal photoproduction is unknown as is its  
dependence on the nuclear target mass.

It is difficult to formulate theoretical implications of these results in
general terms, i.e., without recurring to specific models. 
For hard processes intuitive physical pictures were
proposed in the literature to explain qualitatively the origin of single-spin asymmetries 
(SSAs) and especially of transverse hyperon polarization;
see, e.g., Refs. \cite{Abramov,Burkardt,Hannafious:2008dx}.
The usual factorization of QCD processes into products of distribution functions,
hard amplitudes, and fragmentation functions \cite{Mulders} requires a large momentum in
the hard amplitude. This requirement is not fulfilled for quasireal photoproduction and at the relatively 
small transverse momenta of the present experiment, as opposed to deep-inelastic lepton-nucleon scattering. 
Still, a smooth dependence on the
virtuality $Q$ of the photon would be expected such that a comparison of results from
single-spin asymmetries in leptoproduction 
and in p-p and p-A collisions as well as corresponding theoretical calculations~\cite{Koike:2002ar}
is interesting, especially because even the sign of
theoretical expectations is still under debate; see Ref.~\cite{Kang}.
Recently, the theoretical situation seems to have been clarified by Ref.~\cite{Kanazawa},
which stressed that $p_T$-dependent fragmentation plays an important
role. If so, one might also expect sizable asymmetries in soft $p_T$-dependent fragmentation.

\section{Acknowledgments}

We gratefully acknowledge the {\sc DESY} management for its support and the staff
at {\sc DESY} and the collaborating institutions for their significant effort.
This work was supported by 
the Ministry of Education and Science of Armenia;
the FWO-Flanders and IWT, Belgium;
the Natural Sciences and Engineering Research Council of Canada;
the National Natural Science Foundation of China;
the Alexander von Humboldt Stiftung,
the German Bundesministerium f\"ur Bildung und Forschung (BMBF), and
the Deutsche Forschungsgemeinschaft (DFG);
the Italian Istituto Nazionale di Fisica Nucleare (INFN);
the MEXT, JSPS, and G-COE of Japan;
the Dutch Foundation for Fundamenteel Onderzoek der Materie (FOM);
the Russian Academy of Science and the Russian Federal Agency for 
Science and Innovations;
the Basque Foundation for Science (IKERBASQUE) and the UPV/EHU under program UFI 11/55;
the U.K.~Engineering and Physical Sciences Research Council, 
the Science and Technology Facilities Council,
and the Scottish Universities Physics Alliance;
as well as the U.S.~Department of Energy (DOE) and the National Science Foundation (NSF).

\end{document}